# CO-DESIGNING IN SOCIAL VR

*Process awareness and suitable representations to empower user participation*

TOMÁS DORTA[1], STÉPHANE SAFIN[2], SANA BOUDHRAÂ[3] and EMMANUEL BEAUDRY MARCHAND[4]
[1,3,4]*Design Research Laboratory Hybridlab, University of Montreal*
[1,3,4]*{tomas.dorta|sana.boudhraa| emmanuel.beaudry.marchand}@umontreal.ca*
[2]*Télécom ParisTech i3-CNRS*
[2]*stephane.safin@telecom-paristech.fr*

**Abstract.** To allow non-designers' involvement in design projects new methods are needed. Co-design gives the same opportunity to all the multidisciplinary participants to co-create ideas simultaneously. Nevertheless, current co-design processes involving such users tend to limit their contribution to the proposal of basic design ideas only through brainstorming. The co-design approach needs to be enhanced by a properly suited representational ecosystem supporting active participation and by conscious use of structured verbal exchanges giving awareness of the creative process. In this respect, we developed two social virtual reality co-design systems, and a co-design verbal exchange methodology to favour participants' awareness of the co-creative process. By using such representations and verbal exchanges, participants could co-create with more ease by benefiting from being informed of the process and from the collective immersion, empowering their participation. This paper presents the rationale behind this approach of using Social VR in co-design and the feedback of three co-design workshops.

## 1. Introduction

Design projects are complex because they relate to a wide variety of end-users and also involve multiple stakeholders along the process. Going beyond current participatory design approaches, we claim new methods are needed to allow better involvement of users in these projects, in a perspective of active co-design (bottom-up), instead of merely looking for user approval or consensus on previously set decisions (top-down). But involving non-designers in the hearth of design raises two major issues. On one hand, public organizations and design offices dictate or mediate the engagement of projects' prime stakeholders by using processes, representations, methodologies and technologies that are not well adapted for lay persons, reducing them to the role of receptors instead of actors to whom the final designs belong. On the other hand, spontaneous initiatives coming from users themselves as non-designers need the appropriate means to result in relevant design solutions. Co-design must give the opportunity to all the participants to co-create ideas upstream, simultaneously (co-ideation), in a democratic (horizontal) and multidisciplinary way. Nowadays, co-design processes involving non-designers tend to limit their contribution to proposing general, abstract and "immature" design ideas through creative techniques like brainstorming and post-it notes, without further diving into design concepts at a mature level and resolving them. Consequently, the co-design approach needs to be enhanced by an appropriate representational ecosystem supporting active participant intervention and through the identification of structured verbal exchanges providing *awareness* of the ongoing creative process. To this end, we developed two Social VR (Virtual Reality) co-design systems (see section 6), to support democratic user participation. Thanks to its natural interaction, immersion and embodiment without wearing VR headsets, users can resolve design problems together with other specialist collaborators, using life-sized representations, immersive and 3D sketches, actively interacting with design ideas. Diverse groups of co-located participants (including architects) are immersed within different kinds of representations. Because complexe codified representations like plans or scaled presentation mock-ups are challenging to grasp for neophytes, Social VR supports representations suited for the creative phases: real-scale, immersive, with high evocative power, and with the easiness of sketch-based graphical expression. Also, we developed a co-design methodology to support and bring design process awareness to participants through structured conversations. Pairing such structuring of discourse along with the aforementioned representational ecosystem, supports user co-creation and exchanges with greater ease by being more aware and immersed in the project, thus empowering them through facilitated participation via Social VR.

## 2. Participatory design

Participatory design projects aim for user implication during the process but nevertheless, users rarely have the power to be fully involved due to several reasons. One is that the project's owner often arrives with projects designed beforehand, only seeking for user approval (Jutand 2013), persuading them that they were involved in the creative process while all decisions are already set. Another reason is that users themselves as non-designers are not properly supported to fully engage during the first steps of the design process (ideation) where main, driving ideas are generated and most important decisions are made. Consequently, end-users have problems appropriating or embracing such projects. Risks of integration conflicts within the context could arise (Lehtovuori 2016) and designs could even fail to fulfill their initial goals. Moreover new initiatives are emerging, allowing users to participate firsthand in design processes that concern them, no more only as passive subjects to be consulted, but rather as co-authors of design proposals (e.g. Living Labs). Nonetheless, under the appearance of participation, these initiatives are often limited to a collection of primitive ideas or are intended only to obtain the approval of people. Receivers or users must assume new roles. They should no longer be passive or solely capable of a posteriori reaction but should become more and more considered as referents of the project. There is a need for *informed, conscious awareness* of the creative process (Farina et al. 2014). From being simple users, they have to be more involved as the ones for whom the projects are truly conceived, then becoming active contributors along the process (Jutand 2013). The "author project" is replaced by a networked, pluralistic design, where the user also becomes a co-author and participates in the design. The role of the designer then evolves towards that of a "midwife" or, as Sanders and Stappers (2008) describe, an *accompanist*.

## 3. Co-design approach

To cope with this, new design approaches named co-design or co-creation consist in a creative and collective activity where two or more persons, or a community, get involved along with designers and architects in a design project that matters to them (Sanders and Stappers 2008). Kleinsmann and Valkenburg (2008) describe co-design as a process where actors from various disciplines share their knowledge of the process and the design content in order to create a shared understanding of both aspects and achieve a common goal. Besides, co-design has been proposed as a way to improve idea generation, decision making, communication and creativity (Steen et al. 2011). Steen (2013) emphasizes that co-design focuses on the concrete practices of people, their experiences and the role of practical knowledge. Co-design is also understood as a technique of collaborative design involving participants simultaneously in a co-creation process. The simultaneous character of co-design is an important component in the process compared to cooperative design consisting in an asynchronous process (Kvan 2000). Here, cooperative design refers to when a group of individuals accomplish different tasks separately in order to achieve a common goal (Achten 2002). Participatory design and co-design both share the advantages of integrating individuals' skills and experience(s), as revealed through the unfolding exchanges (Schuler and Namioka 1993). Moles (1986) calls them *repertoire*, that is to say, a constitutive knowledge of each one, which joins Schön's own notion of repertoire (1985). Mattelmäki et al. (2011) raise the question of power attribution to all participants in a fair way (democratic) in collaborative design processes. We define co-design as a particular collaborative design process. It requires engaging the group in simultaneous mutual collaboration and actively involving all participants. The generation of concepts then takes place together through a spontaneous exchange where ideas are built, negotiated and accomplished by a common agreement. To build on these concepts, our proposed co-design approach using Social VR (see section 6) has two main ingredients: a process made aware to the participants through the structure of verbal exchanges, and a particular representational ecosystem that allows layperson to fully understand design proposals and fully engage in the co-design process.

### 4. Process awareness: informed through verbal exchanges

Verbal communication constitutes, above sketches, the prime design tool allowing designers to externalize intentions and launch ideation (Jonson 2005). In a collaborative setting, designers communicate their ideas verbally, but also through gestures and representations (graphical and physical). The conversation around collaborative ideation is a kind of rhetoric that seeks to build consensus where all stakeholders aim to resolve a common problem (Asher and Lascarides 2003). In previous work where we analyzed the design discourse during collaborative ideation between designers, we identified recurrent conversational patterns at different moments of the process (Dorta et al. 2011). These patterns characterize the designers' thinking in action as well as the maturity degree of design ideas. We distinguished six main elements: *naming, constraining, proposing* (key element), *negotiating (questioning and explaining), decision making and moving forward the idea* via representations (sketching and gesturing). The first five relate to verbal exchanges while the latter is an act of representation that transforms the design situation (Goldschmidt 1990; Valkenbug and Dorst 1998). In addition, we also identified four recurrent conversational patterns that are linked to different phases of the co-ideation process and different levels of idea maturity: collaborative conversations (presenting and discussing), immature collaborative ideation loops (immature CI-loop), mature CI-loop and collaborative moving. The two kinds of CI-loops are the most important patterns in the process since they are where ideas are proposed and negotiated. During immature CI-loops participants are focusing on *what* they have to design using longer negotiations, while in mature CI-loops they reflect on *how* the design concept will be anchored to reality (Dorta et al. 2011). Moreover, in 2015 we developed a co-design course at the university level where we teach this patterns identification as a design strategy to students, in order to make them *aware of the creative process*. In addition, we have been realizing multiple local and international co-design workshops (involving mainly non-designers) using this approach to make the co-design process more explicit to further support participation and contribution from all actors.

### 5. Representational ecosystem and Social VR

Involving design lay persons (untrained in design) in collaborative design projects faces a major practical challenge: an adapted representational ecosystem. Non-designers can have trouble visualizing and placing themselves into design ideas. In particular, the representations that are mobilized by design professionals (scaled models, technical drawings, Computer-Aided Design "CAD" models) do not always make it possible to evoke effectively the characteristics of the project (Van de Vreken and Safin 2010). In addition, there is a lack of adequate representational tools for inter-professional actors (designers, engineers, users, etc.) to graphically and physically externalize their ideas, that is, to involve positively users in design and ultimately transform non-specialists into real actors. Indeed, non-designers are rarely trained in freehand sketches, perspective views or 3D modelling, tools that are central in "professional" design practice. Externalizing and framing design ideas are also critical for the construction of a common ground in collective processes, which is necessary for the development of quality projects (Stempfe and Badke-Schaub 2002). Kvan (2000) indicates that collaboration is more complicated than the mere participation of a group of individuals. It requires a greater synchronization of work and simply bringing new software and hardware is not sufficient to create the right environment to foster collaboration (Kvan 2000). Thus, communication plays a decisive role for this synchronization which subsequently allows the evolution of each person's mental representations (Darses 2006).

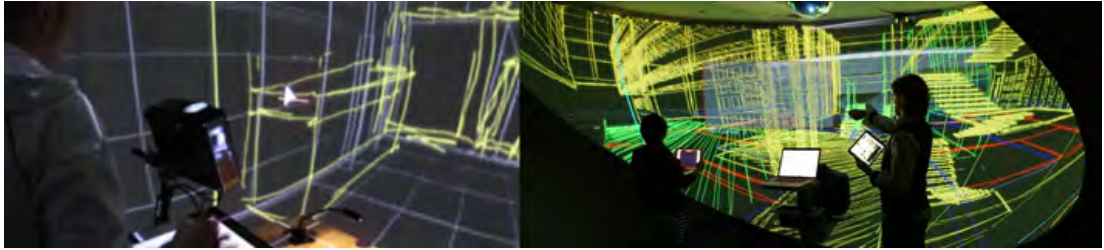

Figure 1. HIS, individual immersive sketching + model making; Hyve-3D, multi-user immersive 3D sketching.

We have previously studied the characteristics of the representational ecosystem that support or enhance the co-design processes (Dorta et al. 2016). Four main required elements were found: it has to (1) be *hybrid*, mixing analog (made by hand) and digital representations; (2) allow project visualization *using several scales* including life-sized; (3) *support active co-design* by allowing all users to represent through sketches and physical models and manipulate these representations, all in an intuitive way; and finally, (4) the ecosystem must grant opportunities to use and *bridge across several kinds of representations*, including physical (models) and graphical (different views, photos, etc.). For this purpose, since 2007 we have been developing immersive co-design platforms HIS (Dorta 2007) and Hyve-3D (Dorta et al. 2016), its successor (Figure1), improving local and remote design communication. These systems allow local immersive co-ideation while integrating themselves to the representational ecosystem. The systems consolidate multiple representations which support co-ideation without interfering with the process, in contrast to traditional CAD tools. It permits to work at different scales (including life-sized) with hybrid representations and to co-ideate in an active way via immersive and 3D sketches, immersive model making (using physical mock-ups) and interaction using handheld devices. Both systems are Social VR systems supporting the communication between local participants because they don't use occlusive headsets that hinder their verbal and non-verbal communication (e.g. design gestures) allowing to share the collective design experience.

## 6. Co-designing in Social VR approach

This paper presents the Co-designing in Social VR approach we put in place on six-hour (9am to 4pm) workshops, a duration which ensure people participation. During the workshops, participants *that are not designers or architects* are able to generate *mature* design proposals in the form of scaled physical models and immersive virtual representations (immersive and 3D sketches on a basic 3D geometry used as context). Designers and architects can also get involved in the project as well as engineers, other specialists, and design students as *design accompanists*. The workshop is structured in different sessions, each focused on a specific project stage: problem framing, immature co-ideation, mature co-ideation and final presentation. During each session, participants alternate between Master classes on core concepts (co-design vs. cooperation, verbal design exchanges and the representational ecosystem) and design studio in the representational ecosystem. The goal is to show the rationale underpinning the design's verbal and representational exchanges, after each step is realized, to favour a better comprehension of the process by the participants. Concerning the representational ecosystem, participants alternate using two kinds of representations. (1) *Traditional graphical and physical representations* consisting mostly in freehand sketching on paper and working with a physical model where all the architectural components (walls, furniture, columns, etc.) are already pre-cut in foam-board or cardboard at the right scale (depending on the dimension of the given project). The model can be easily assembled and manipulated collaboratively by the participants This facilitates project discussion and manipulation through their modification as *boundary objects* (Arias and Fischer 2000). (2) *Social VR system* in order to visualize the project life-sized and sketch immersively or in 3D (Hyve-3D). The mock-up can be brought inside the system's space by using a small 360° camera inside the model and experience it in immersion, from within (HIS). At the end, the physical models and the virtual models on the Social VR system are used to deliberate the best proposition (several if the group of participants is large). The physical and virtual models then also provide final 3D representations of the project used to show to project makers. The schedule of the day consists of four one-hour sessions, in rooms dedicated to each group (around 6 persons each max.). During each session, the groups are called in turn in the Social VR system for 20 minutes to design within the immersive environment. These sessions are interspersed with pauses and theoretical returns on their co-design activities by the facilitators. Each session is focused on a specific aspect of design conversation: the first session on the problem (re-)definition and the collaborative conversations CC; the second session on proposing a large set of ideas (Immature CI loops); the third session on the concept deepening (Mature CI loops); and the last on the preparation of the final presentation. At the end of the day, formal presentations are held, using physical models and hybrid representations in the Social VR system as support. These presentations lead to the selection of the most collegially promising project among the proposed solutions made by each team.

## 7. Examples of Co-designing in Social VR workshops

We realized several co-design workshops using this approach and using different kind of user interactions: *mixed reality*, installing the Social VR system in the same space to be designed; *scale model endoscopy*, where the participants were immersed inside their physical models; and *collective 3D sketches* using the Hyve-3D, in which several users sketch simultaneously in 3D using as contextual background a dynamic 3D model. In this paper we present only those workshops involving non-designers as participants. Each of these projects was focused on preliminaray phase of the (re)design of the interior of an existing space, or a space currently being buit. The first two were executed in the context of a business school of an university campus and the third for a university's library in another country. All involved were participants institutions' employees and users (professors and students that attend that facility). As mentioned before, in these workshops some architects and design students participated as accompanists to facilitate the design process. Among their tasks, the accompanists focused on the proper use of the different kinds of representations, on exploring alternatives to avoid getting stuck with only one idea or on helping in the discussions and the development of design arguments.

## 7.1. MIXED REALITY: BUSINESS SCHOOL'S TEACHING SPACE

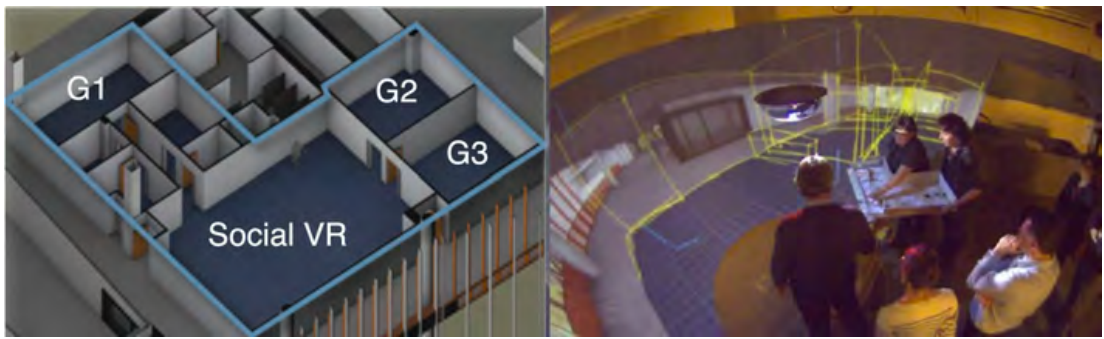

Figure 2. (Right) Space to be redesigned and the sub-spaces of the workshop for each group (Left) Social VR displaying a virtual model and digital sketches combined with a scaled model.

The workshop we conducted consists in the re-design of an existing teaching space in the building of the business school, underused and inappropriate (Figure 2). This space is composed of a large main room and six small rooms and is designed to support negotiation courses. Participants (24) were divided into three groups, and each group had the goal to achieve a design proposal for the future space. Since the Social VR systems are mobile, this workshop took place into the same space to be redesigned, allowing participants to use elements of the real space as a support and as referents for discussion in a context of mixed reality (Safin et al. 2013). Figure 2 shows the different sub-spaces (one room per group) and a central room with the Social VR system (HIS) installed in the center of the redesigned space. Digital models of this space were built and displayed in the HIS as 360° panoramas. Participants therefore had the opportunity to see an immersive projection of a digital model of the design space in the heart of this real space, with the possibility of aligning the same point of view. In the HIS, no special skills was required since the participants were sketching (life-sized) over the immersive panorama. Spontaneously, some participants were manipulating the digital pencil of the Wacom Cintiq tablet of the HIS, while others were designing and discussing alternating from the immersive representation and the real space, visually accessible from the HIS.

7.2. SCALE MODEL ENDOSCOPY: BUSINESS SCHOOL'S LIBRARY

This workshop was related to the redesign of the business school's main library. Several library users and employees also participating in the workshop (25), using the same representational ecosystem and the HIS of the previous workshop. However, this time the activity took place in our laboratory's rooms. They developed several physical models and became immersed inside those scaled model by using the 360° camera (Figure 3).

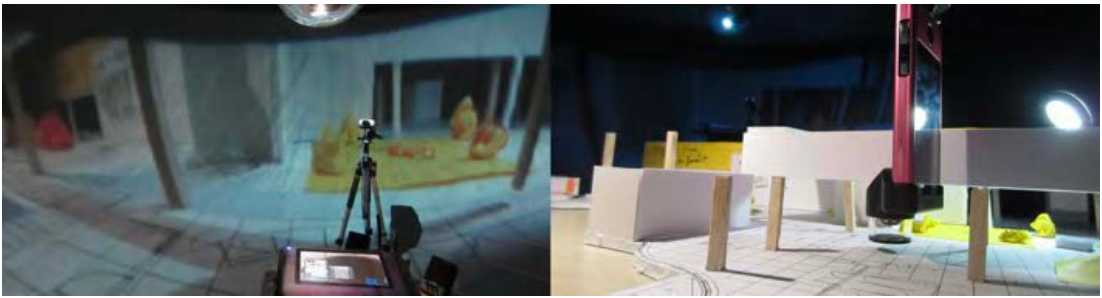

Figure 3. (Left) Life-sized immersive view on HIS of the rough physical model using the 360° camera (Right).

7.3. COLLECTIVE 3D SKETCHING: UNIVERSITY'S LIBRARY

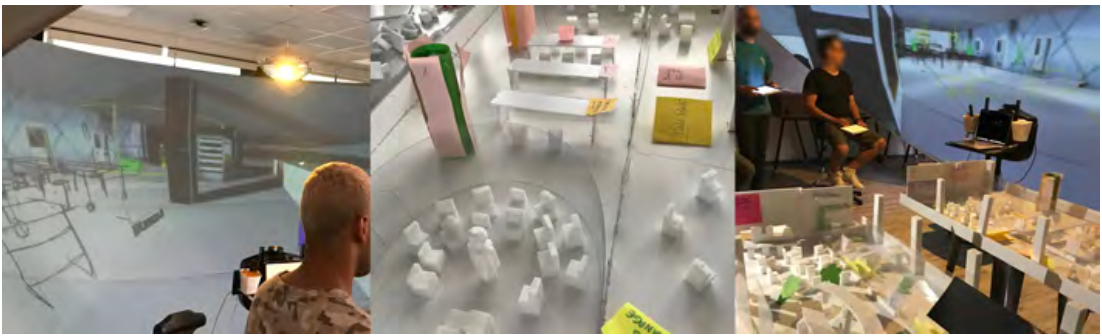

Figure 4. Hyve-3D allowing collective 3D sketching; and rough scaled physical models.

This last example took place in another university (in another country) where, this time the Hyve-3D was transported and installed. The participants were again users and employees of the the library and 1 design accompanists per team (2 teams of 6). Concerning the representational ecosystem, once again scale model pieces were pre-cut and used as boundary objects to assemble the physical model while designing the ideas. The Hyve-3D, compared to HIS, allows collective 3D sketching using two handheld tablets. In that concern, we also provide a basic 3D model as a contextual background, made from SketchUp and textured in a baking process using a rendering software (Figure 4). In two periods of 30 min., four participants were trained to sketching in 3D using the two tablets of the Hyve-3D.

## 8. Feedback from the workshops

All groups came to interesting results, even surprising, given the initial objectives, and according to the client and the participants. All participants reported having experienced a rich and interesting day, and having actually contributed to their project. Although this informal observation can't allow us to draw definitive conclusions, it sheds light on the relevance of this new co-design approach, which is based on the representational ecosystem (Social VR and traditional tools such as pen-and-paper and physical models) and the process awareness through verbal exchanges. Another interesting informal observation relates to the relevance of the phenomenon of Social VR for the construction of a shared vision of the project. Indeed, we found that the different teams right out of Social VR directly address new issues of design, without going through a reformulation of the discussions made in the system. So it seems that the teams are directly operational for the design starting from Social VR, indicating its capacity to support a truly shared vision of the problem. The Social VR and the physical models have been widely used both by all participants, even those who were not familiar with design (business students, teachers, etc.). These representation modes seem more intuitive than the conventional representation tools like 2D or perspective sketches. The directly perceptible scale that these two modes of representation convey seems to be a determinant of the perceived effectiveness. It appears that both the Social VR systems that the physical model were relevant platforms for supporting multidisciplinary collaborative design. These two elements of the representational ecosystem seem to support different ways to design, which are complementary: mock-ups afford assembly of elements, spatial arrangement with a global view of the space, whereas Social VR support the contextual sketching and proporsions of proposed design elements in space (new furniture, walls, etc.). Finally, we made another interesting observation: as the workshops were progressing, we have seen an evolution of the commitment of the participants. While the first sessions were held calmly, with very fluid speech turns, the last sessions were characterized by more assertive behaviours from all the participants: more verbal exchanges, many collective actions on representation tools (mainly physical model) and more engaged postures. Finally, although all stakeholders were very enthusiast about the process and the project, the impact of these workshops on the design project has not been formally assessed.

## 9. Conclusions

This paper presented a new co-design approach based on Social VR combined with physical mock-ups as boundary objects. Also, verbal exchanges structuration method allows non-designers participants to get informed and be aware of the creative process. Feedback from real co-design workshops involving lay persons as participants was reported. Different kinds of interaction were exposed: mixed reality, endoscopy of physical models and collective immersive 3D sketching. The approach allows during a single day workshop to propose mature designs via 3D representations to engage with real project makers. Even if specific Social VR devices were used, other VR systems as headsets combined with big projections and immersive walls could be used. The goal is to consider the use of the appropriate representational ecosystem and structured verbal exchanges to allow awareness and improve participations of non-designers in co-design processes.

**References**

Achten, H.: 2002, Requirements for collaborative design in architecture, *6th Design & Decision Support Systems in Architecture & Urban Planning Conference*, Eindhoven, 1-13.
Arias, E. G. and Fischer, G.: 2000, Boundary objects: Their role in articulating the task at hand and making information relevant to it, *International ICSC Symposium on Interactive & Collaborative Computing*, Australia, 567–574.
Asher, N. and Lascarides, A.: 2003, *Logics of conversation*, Cambridge University Press.
Darses, F.: 2006, Analyse du processus d'argumentation dans une situation de reconception collective d'outillages, *Le travail humain*, **69**(4), 317-347.
Dorta, T.: 2007, Implementing and Assessing the Hybrid Ideation Space: a Cognitive Artifact for Conceptual Design, *Int. Journal of Design Sciences and Technology*, **14**(2), 119-133.
Dorta, T., Kalay, Y. E., Lesage, A. and Pérez, E.: 2011, Elements of Design Conversation in the interconnected HIS, *Int. Journal of Design Sciences and Technology*, **18**(2), 65-80.
Dorta, T., Kinayoglu, G. and Hoffmann, M.: 2016, Hyve-3D and the 3D Cursor: Architectural co-design with freedom in Virtual Reality, *International Journal of Architectural Computing*, **14**(2), 87-102.
Farina, C., Kong, H., Blake, C., Newhart, M. and Luka, N.: 2014, Democratic deliberation in the wild: The McGill Online Design Studio and the RegulationRoom Project, *Fordham Urban Law Journal*, **41**(5), 1527-1580.
Goldschmidt, G.: 1990, Linkography: assessing design productivity, *Cyberbetics and System'90, Proceedings of the Tenth European Meeting on Cybernetics and Systems Research*, 291-298.
Jonson, B.: 2005, Design ideation: the conceptual sketch in the digital age, *Design Studies*, **26**(6), 613-624.
Jutand, F. and Dartiguepeyrou, C.: 2013, *La métamorphose numérique: vers une société de la connaissance et de la coopération*, Alternatives, Paris.
Kleinsmann, M. and Valkenburg, R.: 2008, Barriers and enablers for creating shared understanding in co-design projects, *Design Studies*, **29**(4), 369-386.
Kvan, T.: 2000, Collaborative design: what is it?, *Automation in construction*, **9**(4), 409- 415.
Lehtovuori, P.: 2016, *Experience and conflict: The production of urban space*, Routledge.
Mattelmäki, T. and Sleeswijk Visser, F.: 2011, Lost in Co-X: Interpretations of Co-design and Co-creation, *Diversity and Unity, Proceedings of IASDR2011*.
Safin, S., Lesage, A., Hébert, A., Kinayoglu, G. and Dorta, T.: 2013, Analyse de référents de l'activité de co-design dans un contexte de Réalité Mixte, *Proceedings of Interaction Homme-Machine, IHM*, 23-32.
Sanders, E. B.-N. and Stappers, P. J.: 2008, Co-creation and the new landscapes of design, *Co-Design*, **4**(1), 5-18.
Schuler, D. and Namioka, A.: 1993, *Participatory design: Principles and practices*, CRC Press.
Steen, M.: 2011, Tensions in human-centred design, *CoDesign: International Journal of CoCreation in Design and the Arts*, **7**(1), 45-60.
Steen, M.: 2013, Co-design as a process of joint inquiry and imagination, *Design Issues*, **29**(2), 16-28.
Stempfle, J. and Badke-Schaub, P.: 2002, Thinking in design teams: an analysis of team communication, *Design Studies*, **23**, 473-496.
Valkenburg, R. and Dorst, K.: 1998, The reflective practice of design teams, *Design Studies*, **19**(3), 249-271.
Van de Vreken, A. and Safin, S.: 2010, Influence du type de représentation visuelle sur l'évaluation de l'ambiance d'un espace architectural, *Proceedings of Interaction Homme-Machine, IHM*, 49-56.